\documentclass[aps,showpacs,twocolumn]{revtex4}
\usepackage{graphicx}
\usepackage[T1]{fontenc} 
\usepackage{epsfig,latexsym}
\usepackage{color}
\usepackage[latin1]{inputenc}
\usepackage{amsmath}
\usepackage{amssymb}
\usepackage{amsfonts}
\usepackage{indentfirst}
\usepackage{ae}
\usepackage{float}

\DeclareMathOperator{\Tr}{Tr}

\begin{document}

\title{\bf Entanglement between two scalar fields in an expanding spacetime}

\author{Helder Alexander $^a$}\email[]{helder@fisica.ufmg.br}
\author{I. G. da Paz $^b$}\email[]{irismarpaz@ufpi.edu.br}
\author{Marcos Sampaio $^a$}\email[]{msampaio@fisica.ufmg.br}

\affiliation{$^a$ Universidade Federal de Minas Gerais - Departamento de F\'{\i}sica - ICEX \\ P.O. BOX 702, 30.161-970, Belo Horizonte MG - Brazil}
\affiliation{$^b$ Departamento de F\'{\i}sica, Universidade Federal do Piau\'{\i}, Campus Ministro Petr\^{o}nio Portela, CEP 64049-550, Teresina, PI, Brazil}

\begin{abstract}

We study the evolution of the two scalar fields entangled via a
mutual interaction in an expanding spacetime. We compute the
logarithmic negativity to leading order in perturbation theory and
show that for lowest order in the coupling constants, the mutual
interaction will give rise to the survival of the quantum
correlations in the limit of the smooth expansion. The results
suggest that interacting fields can codify more information about
the underlying expansion spacetime and lead to interesting
observable effects.

\end{abstract}

\maketitle

\paragraph*{\bf Introduction} - Entangled states as firstly described
in the seminal paper of Einstein, Podolsky and Rosen \cite{EPR}
has been the subject of many studies, principally due to the fact
that they has emerged as indispensable physical resource for the
performance of present-day quantum information tasks, such as
quantum communication \cite{BennettI}, quantum teleportation
\cite{BennettII}, quantum cryptography \cite{Ekert}, superdense
coding \cite{Horodecki} and quantum computation \cite{Steane}.
Recently, much attention has been directed to understanding how
these correlations behave in a relativistic setting \cite{Wang}.
From the practical viewpoint a good example is Relativistic Quantum
Metrology, which exploits non-inertial effects on quantum
entanglement to develop extremely high-precision parameter
estimation protocols, with signal-to-noise ratios that may achieve
the Heisenberg limit \cite{Heisenberg}. These metrology protocols
have been employed to conceive precision measurements of Unruh
temperatures and effects of gravity on entanglement
\cite{Unruh1}\cite{GBE} \cite{Downes}, as well as to conceive novel
schemes for gravitational wave detection which may provide feasible
alternatives to experiments such as LIGO ( Laser Interferometer
Gravitational-Wave Observatory) and may be within technological
reach in the near future \cite{GW, Rideout}.

However, the interest on relativistic effects on entanglement does
not arise only from its role as a resource for quantum information
tasks. Sometimes entanglement itself may actually encode the
parameters of interest in relativistic settings. One example is
given by the parameters of the large-scale spacetime metric. It has
long been known that in the context of expanding spacetimes in
general, pairs of entangled particles are dynamically created into
modes of opposite momenta of a free scalar field \cite{Birrell} for
instance. The amount of entanglement generated by a period of
expansion may be determined by the spacetime metric. Therefore
\cite{fuentes4}, \cite{fuentes3}, and \cite{Wang2} suggest that
measurement of these quantum correlations offers a tool to estimate
the parameters that characterize the scale factor. Moreover in
\cite{Montero, Martinez} it is discussed  that cosmic neutrinos may
encode entanglement generated in the early universe epoch which
could possibly survive to be detected, since they interact very
weakly with other sources of energy and matter.

The role of interactions is evidently important in this
proof-of-principle level as new phenomena occur, such as a
competition between multiparticle production from the vacuum and
thermalization \cite{Kodama, BLHu}. In this context the interaction
leads the system towards equilibrium, while the spacetime expansion
deviate the system from equilibrium because of entropy production
and particle creation. Interacting processes over this type of
spacetime background can lead, depending on statistics, either to
gravitational amplification or attenuation of particle creation
\cite{Audretsch}, and the exact impact these effects will have on
the amount of entropy and quantum correlations generated has many
subtleties \cite{Kandrup, LinHu}. Moreover, in the context of
inflationary theory \cite{Star,Pi}, there are several works which
point out how the different aspects of the quantum-to-classical
transition of quantum inflaton fluctuations are realized and favored
when the inflaton participates in interacting processes in general
\cite{Kiefer_2, Kiefer_Polarski, Mazur_Heyl}. Therefore, the
ubiquitous interactions between fields could possibly supress the
modewise entanglement initially present at one of them. If strong
enough, they could render it impossible to use as suggested above.

Thus, it is interesting to quantify and understand the effect of
self-interactions and interactions between quantum fields on
particle creation, entropy generation and quantum entanglement
during a period of spacetime expansion. Of course, treating
interactions in quantum field theory over expanding spacetimes faces
several technical difficulties, which tend to obscure the analysis
of basic qualitative features of quantum information measures.

In this short letter we study a simple toy model with two scalar
fields mutually interacting and generating bipartite correlations in
an expanding spacetime. In particular, we investigate the effects
that an expanding spacetime has on the interaction between the modes
of a massless scalar field $\phi$ and of a massive scalar field
$\psi$. We evaluate the logarithmic negativity to leading order in
perturbation theory and investigate the expansion effect on the
entanglement between the fields $\phi$ and $\psi$. Our results
suggests a possible competition between entanglement production by
interaction and thermalization generated by expansion. This means
that in the regime of smooth expansion $\frac{\rho}{\omega_{k}} \ll
1$ the interaction provide an important contribution to the survival
of the quantum correlations in the distant future.

\paragraph*{\bf The $\phi\psi$ model} - Let us consider two real scalar fields $\phi$ and $\psi$
in a spatially flat Robertson-Walker spacetime with metric
\begin{align}
ds^2 = a^2(\eta)\left(d\eta ^2 - dx^2\right), \label{metric}
\end{align}
where $a(\eta)$ is the scaler factor and $\eta =
\int\frac{dt}{a(t)}$ is the conformal time ranging from $-\infty$ to
$\infty$. The action of the system reads
\begin{align}
S = \frac{1}{2}\int
d^4x\sqrt{-g}[\partial_{\mu}\phi\partial^{\mu}\phi +
\partial_{\mu}\psi\partial^{\mu}\psi + m^2\psi^2 +
2\lambda\phi\psi], \nonumber
\end{align}
where $g$ is the determinant of the metric tensor $g_{\mu\nu}$ and
$\lambda$ is the coupling parameter normalized such that, $|\lambda|
\ll 1$. The dynamics of the fields $\phi$ and $\psi$ in the
interaction picture are governed by the covariant from Klein-Gordon
equations in the curved spacetime
\begin{align}
  \square\phi(\eta, x) &= 0, \label{Eqphi} \\
  (\square + m^2)\psi(\eta, x) &= 0, \label{Eqpsi}
\end{align}
and the state vector $|\Psi\rangle$ of the system satisfies the
Schrodinger's equation
\begin{align}
H_{I}|\Psi\rangle = i\partial_{\eta}|\Psi\rangle ,
\end{align}
where $H_{I}$ is the normal ordered interaction Hamiltonian
\begin{align}
H_{I} = \lambda\int dx\sqrt{-g}\phi(\eta, x)\psi(\eta, x).
\end{align}

The canonical quantization of the fields $\phi$ and $\psi$ are
identical to that in the free field case. Thus one has that
  \begin{align}
  \phi(\eta, x) = \int dk(a_{k}\phi_{k} + a_{k}^{\dagger}\phi_{k}^{*}), \\
  \psi(\eta, x) = \int dk(b_{k}\psi_{k} + b_{k}^{\dagger}\psi_{k}^{*}),
\end{align}
  with
  \begin{align}
  \phi_{k}(\eta, x) = \frac{e^{ikx}}{\sqrt{2\pi}}u_{k}(\eta), \\
  \psi_{k}(\eta, x) = \frac{e^{ikx}}{\sqrt{2\pi}}v_{k}(\eta),
\end{align}
where the operators $a_{k}$, $a_{k}^{\dagger}$, $b_{k}$, and
$b_{k}^{\dagger}$ satisfy the usual commutation relations
$[a_{k},a^{\dagger}_{k'}] = [b_{k},b^{\dagger}_{k'}] =
\delta_{k,k'}$, and $u_{k}$ and $v_{k}$ are solutions of the
equations
\begin{align}
  u''_{k}(\eta) + k^2u_{k}(\eta) = 0, \label{Equk}\\
  v''_{k}(\eta) + (k^2 + a^2(\eta)m^2)v_{k}(\eta) = 0. \label{Eqvk}
\end{align}

\begin{figure}[htp]
    \centering
    \includegraphics[scale=0.5]{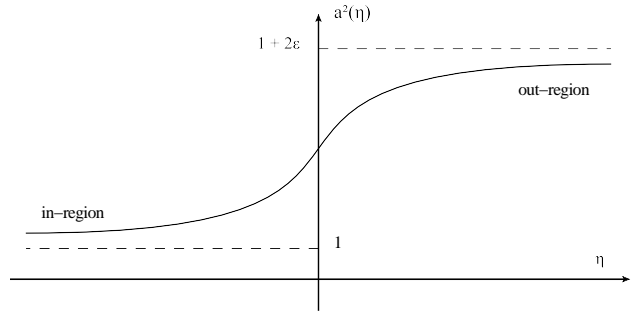}
    \caption{Conformal factor for a toy model universe which possess asymptotic regions.}
    \label{fig0}
\end{figure}

We suppose that $a^2(\eta) = 1 + \epsilon(1 + \tanh(\rho\eta))$,
where $\epsilon$ and $\rho$ controlling the volume and rapidity of
the expansion, respectively. The spacetime becomes flat since
$a^2(\eta)$ is sufficiently smooth and approaches constant values in
the distant past $a^2(\eta \rightarrow -\infty) = 1$ and far future
$a^2(\eta \rightarrow \infty) = 1 + 2\epsilon$,   as illustrated in
figure (\ref{fig0}). In such asymptotic regions Poincar\'e
invariance guarantees the existence of a time-like Killing vector
field $\partial_{\eta}$ orthogonal to all spacelike hypersurfaces of
constant conformal time, and therefore there is an unambiguous way
to distinguish positive- and negative-frequency modes solution of
the field equations (\ref{Equk}) and (\ref{Eqvk}).

In this scenario in which spacetime possesses stationary asymptotic
regions, the solutions of the equation (\ref{Equk}) in the
asymptotic regions are equivalent, i.e.,
\begin{align}
u_{k}^{\mathrm{in}} = u_{k}^{\mathrm{out}} =
\frac{e^{-ik\eta}}{\sqrt{2k}}. \label{uk}
\end{align}
Notice that in the particular case of two spacetime dimensions the
theory of massless scalar field is conformally invariant, and as a
consequence no particles are present in the asymptotic future. On
the other hand, the solutions of the equation (\ref{Eqvk}) in the
asymptotic regions are
\begin{align}
v_{k}^{\mathrm{in}} &= \frac{e^{-i\omega_{k}^{\mathrm{in}}\eta}}{\sqrt{2\omega_{k}^{\mathrm{in}}}}, \quad \text{in-region}, \label{vk}\\
v_{k}^{\mathrm{out}} &=
\frac{e^{-i\omega_{k}^{\mathrm{out}}\eta}}{\sqrt{2\omega_{k}^{\mathrm{out}}}},
\quad \text{out-region},
\end{align}
where $\omega_{k}^{\mathrm{in}} = \sqrt{k^2 + m^2}$ and
$\omega_{k}^{\mathrm{out}} = \sqrt{k^2 + (1 + 2\epsilon)m^2}$. These
asymptotic solutions are connected by a Bogoliubov transformation
that only mixes modes of the same momentum $k$:
\begin{align}
v_{k}^{\mathrm{in}}(\eta) = \alpha_{k}v_{k}^{\mathrm{out}}(\eta) +
\beta_{-k}v_{k}^{\mathrm{out}*}(\eta),
\end{align}
where the Bogoliubov coefficients $\alpha_{k}$ and $\beta_{k}$
satisfy the normalization condition $|\alpha_{k}|^2 - |\beta_{k}|^2
= 1$. In this particular case they are readily evaluated to
\begin{align}
\alpha_{k} &= \sqrt{\frac{\omega_{k}^{\mathrm{out}}}{\omega_{k}^{\mathrm{in}}}}\frac{\Gamma(1 - \frac{i\omega_{k}^{\mathrm{in}}}{\rho})\Gamma(-\frac{i\omega_{k}^{\mathrm{out}}}{\rho})}{\Gamma(-\frac{i\omega_{+}}{\rho})\Gamma(1 - \frac{i\omega_{+}}{\rho})}, \nonumber \\
\beta_{k} &=
\sqrt{\frac{\omega_{k}^{\mathrm{out}}}{\omega_{k}^{\mathrm{in}}}}\frac{\Gamma(1
-
\frac{i\omega_{k}^{\mathrm{in}}}{\rho})\Gamma(\frac{i\omega_{k}^{\mathrm{out}}}{\rho})}{\Gamma(\frac{i\omega_{-}}{\rho})\Gamma(1
+ \frac{i\omega_{-}}{\rho})}, \label{BC}
\end{align}
where $\omega_{\pm} = \frac{1}{2}(\omega_{k}^{\mathrm{out}} \pm
\omega_{k}^{\mathrm{in}})$. In the interaction picture, the
Bogolyubov coefficients carry information only about noninteracting
contribution to the total particle creation.

\begin{figure}[htp]
    \centering
    \includegraphics[scale=0.5]{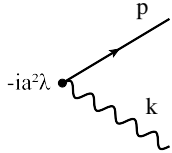}
    \caption{Particle creation out of the vacuum by mutual interaction.}
    \label{fig1}
\end{figure}

\paragraph*{\bf Entanglement state} - Now, let us assume that the composite system
$\phi$ and $\psi$ in the distant past, is a global vacuum (initial condition) for
a given mode $|0_{k}^{\phi}; 0_{p}^{\psi}\rangle^{\mathrm{in}}$. This vacuum state
is annihilated by both $a_{k}$ and $b_{p}$.We denote the  $\phi$ vacuum
by $|0_{k}^{\phi}\rangle^{\mathrm{in}}$ and $\psi$ vacuum by $|0_{p}^{\psi}\rangle^{\mathrm{in}}$ .
Thus the particle creation due to the mutual interaction can be calculated by evaluating the $S$-matrix
to leading order in the interaction picture. Thus to lowest order in $\lambda$,
\begin{align}
S = 1 - i\int_{-\infty}^{\infty}H_{I}d\eta = 1 - i\lambda\int
d^2x\sqrt{-g}\phi\psi.
\end{align}
To this order such an interaction  produces particles in pairs as
depicted in figure (\ref{fig1}), where the probability amplitude is
given by
\begin{align}
^{\mathrm{in}}\langle 1_{k}^{\phi}; 1_{p}^{\psi}|S|0_{k}; 0_{p}\rangle^{\mathrm{in}} &= -2!i\lambda\int d^2x\sqrt{-g}\phi_{k}^{*}\psi_{p}^{*}, \nonumber \\
&= \lambda\delta(k + p)A(k, p),
\end{align}
where
\begin{align}
A(k, p) = \frac{-i}{\pi}\int_{-\infty}^{\infty}d\eta
a^2(\eta)u_{k}^{*\mathrm{in}}(\eta)v_{p}^{*\mathrm{in}}(\eta).
  \end{align}

Inserting the asymptotic mode functions  (\ref{uk}) and (\ref{vk})
in $A(k, p)$, we obtain \cite{Helder}
\begin{align}
A(k, p) &= \frac{-i}{2\pi\sqrt{k\omega_{p}}}\int_{-\infty}^{\infty}d\eta a^2(\eta)e^{-i(k + \omega_{p})}, \nonumber \\
&=  \frac{\pi(1+\epsilon)\delta(k + \omega_{p})}{\sqrt{k\omega_{p}}} \nonumber \\
&+ \frac{\epsilon}{\sqrt{2k\omega_{p}}\rho}\frac{1}{k +
\omega_{p}}\frac{\pi}{\sinh(\frac{\pi}{2\rho}(k + \omega_{p}))}.
\end{align}
The first term proportional to the delta function gives no
contribution to  pair creation process, as it amounts to a shift in
the scale factor \cite{Birrell}. Thus we find
\begin{align}
A(k, p) = \frac{\epsilon}{\sqrt{2k\omega_{p}}\rho}\frac{1}{k +
\omega_{p}}\frac{\pi}{\sinh(\frac{\pi}{2\rho}(k + \omega_{p}))}.
\end{align}
This term expresses a thermal like profile of the vacuum generated
by the interaction over the spacetime evolution.

The initial vacuum state of the total system $|0_{k}^{\phi};
0_{p}^{\psi}\rangle^{\mathrm{in}}$ to lowest order in $\lambda$
reads
\begin{align}
|\Psi\rangle &= N[|0_{k}^{\phi}; 0_{p}^{\psi}\rangle^{\mathrm{in}} \nonumber \\
&+ \frac{1}{2!}\int dkdp ^{\mathrm{in}}\langle 1_{k}^{\phi};
1_{p}^{\psi}|S|0_{k}; 0_{p}\rangle^{\mathrm{in}} |1_{k}^{\phi};
1_{p}^{\psi}\rangle^{\mathrm{in}} + ...], \label{state}
\end{align}
where $2!$ is the symmetry factor and $N$ is the normalization
factor
\begin{align}
N^{-2} &= 1 + \frac{1}{2!}\int dkdp|^{\mathrm{in}}\langle
1_{k}^{\phi}; 1_{p}^{\psi}|S|0_{k}; 0_{p}\rangle^{\mathrm{in}}|^2.
\end{align}
Note that (\ref{state}) is a bona fide entangled state described by
the Hilbert space $\mathcal{H} =
\mathcal{H}_{\phi}\otimes\mathcal{H}_{\psi}$. This means that the
mutual interaction generates bipartite quantum correlations between
the fields $\phi$ and $\psi$.

Consider that the vacuum state $|0_{p}^{\psi}\rangle^{\mathrm{in}}$
in the asymptotic past correspond to a two-mode squeezed state from
the point of view of an inertial observer in the asymptotic future
\begin{align}
|0_{p}^{\psi}\rangle^{\mathrm{in}} = \sqrt{1 - \gamma_{p}}\sum_{n =
0}^{\infty}\gamma_{p}^n|n_{p}^{\psi}, n_{-p}^{\psi}\rangle ,
\label{0p}
\end{align}
where $\gamma_{p} = \left|\frac{\beta_{p}}{\alpha_{p}}\right|^2$.
Since that we are working a single mode, we will drop the frequency
index $k$.

Similarly, the one-particle excitation in the in-vacuum
$|1_{p}^{\psi}\rangle^{\mathrm{in}}$ evolves as
\begin{align}
|1_{p}^{\psi}\rangle^{\mathrm{in}} = (1 - \gamma_{p})\sum_{n =
0}^{\infty}\gamma_{p}^n\sqrt{n + 1}|n + 1_{p}^{\psi},
n_{-p}^{\psi}\rangle . \label{1p}
\end{align}
However, note that for the field $\phi$,
$|0_{k}^{\phi}\rangle^{\mathrm{in}} = |0_{k}^{\phi}\rangle$.

Using the equations (\ref{0p}) and (\ref{1p}), we can rewrite the
equation (\ref{state}) in terms of out-region Fock states for the
field $\psi$
\begin{align}
|\Psi\rangle &= N[ \sqrt{1 - \gamma_{p}}\sum_{n =
0}^{\infty}\gamma_{p}^n|0_{k}^{\phi}; n_{p}^{\psi},
n_{-p}^{\psi}\rangle + \lambda\int dp A(p,-p) \nonumber \\ &\times(1
- \gamma_{p})\sum_{n = 0}^{\infty}\gamma_{p}^n\sqrt{n +
1}|1_{k}^{\phi}; n+1_{p}^{\psi}, n_{-p}^{\psi}\rangle + ...],
\nonumber
\end{align}
This state enable us to construct the density matrix of whole
tripartite state $\hat{\rho}_{k,p,-p}^{\phi\psi} =
|\Psi\rangle\langle\Psi|$ which includes modes of  the two  fields.
Since an inertial observer in the out-region has no access to modes
$-p$, the state $\hat{\rho}_{k,p,-p}^{\phi\psi}$ will be projected
into a mixed state by tracing over all states with modes $-p$
\begin{align}
\hat{\rho}_{k}^{\phi\psi} = \Tr_{-p}[\hat{\rho}_{k,p,-p}^{\phi\psi}]
= (1 - \gamma_{k})\sum_{n =
0}^{\infty}\gamma_{k}^n\hat{\rho}^{\phi\psi}_{n},
\end{align}
where
\begin{align}
\hat{\rho}^{\phi\psi}_{n} &= |0_{k}^{\phi}; n_{k}^{\psi}\rangle\langle 0_{k}^{\phi}; n_{k}^{\psi}| \nonumber \\
&+  \lambda A(k)\sqrt{1 - \gamma_{k}}\sqrt{n + 1}|1_{k}^{\phi}; n+1_{k}^{\psi}\rangle\langle 0_{k}^{\phi}; n_{k}^{\psi}| \nonumber \\
&+ \lambda A^{*}(k)\sqrt{1 - \gamma_{k}}\sqrt{n + 1}|0_{k}^{\phi}; n_{k}^{\psi}\rangle\langle 1_{k}^{\phi}; n+1_{k}^{\psi}| \nonumber \\
&+ \lambda^2|A(k)|^2(1 - \gamma_{k})(n + 1)|1_{k}^{\phi};
n+1_{k}^{\psi}\rangle\langle 1_{k}^{\phi}; n+1_{k}^{\psi}|
\nonumber \\
&+ ... \label{rhon}
\end{align}
\paragraph*{\bf Logarithmic negativity} - The most adequate separability criterion to
estimate the quantum correlation in mixed quantum states is the partial transpose
criterion of Peres-Horodecki \cite{Peres, Horodecki2}. This criterion state that,
if a density matrix is entangled, then its partial transpose has some negative
eigenvalues and hence lacks the positivity required by all density matrix.
It follows that the positivity of the partial transpose (PPT) is a necessary
condition for system, specifically bipartite systems of dimensionality
$2\times 2$ and $2\times 3$. In higher dimensional systems it has been shown
in \cite{Horodecki2} that there are entangled states with positive partial transpose.
These states are known as bound entangled states. Thus, to quantify entanglement we use the logarithmic negativity defined as:
\begin{align}
E_{\mathcal{N}} = \log_2[1 + 2\mathcal{N}],
\end{align}
where $\mathcal{N} = \mathrm{max}\lbrace 0, -\sum_{j}\nu_{j}\rbrace$
and $\nu_{j}'s$ are the negative eigenvalues of the partial
transpose density matrix. Thus, the negative eigenvalues qualifies
and quantifies the entanglement of quantum in a mixed state. It has
been proved that this quantity exhibits monotonic behavior under
Local Operation and Classical Communication (LOCC) operations or
operators that conserves PPT \cite{Plenio}.

\begin{figure}[htp]
\centering
\includegraphics[scale=0.25]{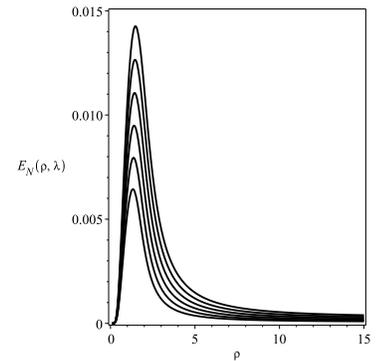}
\caption{Logarithmic negativity as function of the $\rho$  for
different coupling constants $0.0005 \leq  \lambda < 0.001$ with $k
= m = 1$ and $\epsilon = 40$, where higher spectral peaks correspond
to strong couplings.} \label{fig2}
\end{figure}

The partial transpose of (\ref{rhon}) is obtained by exchanging
$|n_{k}^{\phi};n_{k}^{\psi}\rangle\langle m_{k}^{\phi};
m_{k}^{\psi}| \rightarrow |m_{k}^{\phi};n_{k}^{\psi}\rangle\langle
n_{k}^{\phi}; m_{k}^{\psi}|$
\begin{align}
\hat{\rho}_{k}^{\phi\psi\mathrm{T}} = (1 - \gamma_{k})\sum_{n =
0}^{\infty}\gamma_{k}^n\hat{\rho}^{\phi\psi\mathrm{T}}_{n},
\label{rhoT}
\end{align}
with
\begin{align}
\hat{\rho}^{\phi\psi\mathrm{T}}_{n} &= |0_{k}^{\phi}; n_{k}^{\psi}\rangle\langle 0_{k}^{\phi}; n_{k}^{\psi}| \nonumber \\
&+  \lambda A(k)\sqrt{1 - \gamma_{k}}\sqrt{n + 1}|0_{k}^{\phi}; n+1_{k}^{\psi}\rangle\langle 1_{k}^{\phi}; n_{k}^{\psi}| \nonumber \\
&+ \lambda A^{*}(k)\sqrt{1 - \gamma_{k}}\sqrt{n + 1}|1_{k}^{\phi}; n_{k}^{\psi}\rangle\langle 0_{k}^{\phi}; n+1_{k}^{\psi}| \nonumber \\
&+ \lambda^2|A(k)|^2(1 - \gamma_{k})(n + 1)|1_{k}^{\phi};
n+1_{k}^{\psi}\rangle\langle 1_{k}^{\phi}; n+1_{k}^{\psi}|
\nonumber \\
&+ ...
\end{align}
This matrix is infinite dimensional, however it has a block-diagonal
structure which allows us to calculate the eigenvalues analytically
block by block. Note that the eigenvalues corresponding to the first
and last diagonal entries of the matrix are always positive.
Therefore, we must simply diagonalize the matrix
\begin{align}
\hat{\rho}^{\phi\psi\mathrm{T}}_{n} = \left(\begin{matrix}
\lambda^2|A(k)|^2\frac{(1 - \gamma_{k})n}{\gamma_{k}} & \lambda A(k)\sqrt{1 - \gamma_{k}}\sqrt{n + 1} \\
\lambda A^{*}(k)\sqrt{1 - \gamma_{k}}\sqrt{n + 1} & \gamma_{k}
\end{matrix}\right) \nonumber \label{2x2rho}
\end{align}
It follows that the eigenvalues of the density matrix
$\hat{\rho}^{\phi\psi\mathrm{T}}_{n}$ in the $(n, n+1)$ sector are
\begin{align}
\nu_{\pm} = \frac{(1 -
\gamma_{k})\gamma_{k}^n}{2}\left[\frac{\lambda^2A^2(k)n(1 -
\gamma_{k})}{\gamma_{k}} \pm \sqrt{Z_n}\right],
\end{align}
with $$Z_{n} = \left(\frac{\lambda^2A^2(k)n(1 -
\gamma_{k})}{\gamma_{k}}\right)^2 + 4\lambda^2A^2(k)(1 -
\gamma_{k}).$$ Note that the eigenvalues depend on the values of
$\lambda$, $\epsilon$ and $\rho$. In particular, for $\lambda$,
$\epsilon$ and $\rho$ finites, one of the eigenvalues is always
negative. Only in the limit $\epsilon, \rho \rightarrow \infty$
could the negative eigenvalue vanishing ($\nu \rightarrow 0$). It
follows that the logarithmic negativity is given by
\begin{align}
E_{\mathcal{N}} = \log_2[1 + \lambda^2A^2 + \gamma_{k} +
\sum_{n=0}^{\infty}(1 - \gamma_{k})\gamma_{k}^{n}\sqrt{Z_{n}} ].
\end{align}
By a numerical analysis of this expression, summarized in figure
(\ref{fig2}), our first observation is that a degradation of the
entanglement generated by the interaction occurs during the period
of expansion. Figure (\ref{fig2}) shows that if $\lambda$ is fixed,
the degree of entanglement is reduced as the parameter $\rho$
increases. On the other hand, we observe that for small values of
$\rho$ there is an enhancement in the amount of the quantum
correlation as $\lambda$ increases. This is due to the fact that the
interaction destroys the conformal symmetry of the theory.

Notice that the effects of degradation, counteracting entanglement
production by interaction, is dominant in the fast expansion regime,
so that entanglement sudden death is expected in the distant future.
However, in the limit of smooth expansion $\frac{\rho}{\omega_{k}}
\ll 1$ the quantum correlation between the quantum fields in the
early universe could  survive up to the distant future despite
decoherence effects due to interactions. This means that in
principle these quantum correlations are robust enough to be
detectable. Since recent researches has discussed that an initially
entangled state between two free massive scalar fields in de Sitter
space  might affect cosmological observables, such as the power
spectrum and other correlation functions of the inflaton
\cite{Holman, Kanno}.

\paragraph*{\bf Conclusion} - In summary, we have studied a simple toy model of two scalar fields
interacting  in an expanding spacetime. We applied a $S$-matrix
scheme in the interaction picture to investigate the effect of the
dynamics of  spacetime  expansion in quantum entanglement generated
by mutual interaction. In addition, we computed the logarithmic
negativity to leading order in the coupling constant $\lambda$. Our
results show that an increase in the expansion parameter produces a
decreasing  in the quantum entanglement between two scalar fields
whereas increasing the coupling constant within the limit of
perturbation theory enhances quantum entanglement.

These results suggest that during the period of cosmic expansion,
the interaction  is  important to the survival of the quantum
correlations. More realistic extensions of the ideas explored here
may lead to interesting observable effects. This is interesting,
since  entanglement and quantum coherence are affected by the
dynamics spacetime. Another important aspect of this problem that
deserves further study is related to the nature of the interaction.
One possible avenue for further research along this line is to study
the effect of other types of interactions, for instance, weak
interaction responsible by radioactive decay, Yukawa interaction
($g\phi\psi\overline{\psi}$), pion-proton scattering
($g\phi^2\psi\overline{\psi}$), electromagnetic interaction, and a
number of other decay process.

 $$\ast \ast \ast$$







HA and MS acknowledge to the financial support from CNPq (Brazil).

\end{document}